\documentclass[12pt]{article} 
\usepackage{epsfig,latexsym,amssymb}
\title{HOLOGRAPHY, GAUGE-GRAVITY CONNECTION AND BLACK HOLE ENTROPY}
\author{Parthasarathi Majumdar\footnote{$^*$E-mail:
parthasarathi.majumdar@saha.ac.in} \\ Theory Group, Saha Institute of Nuclear Physics, \\Kolkata
700064, India}
\begin{document}
\maketitle

\begin{abstract}
The issues of holography and possible links with gauge theories in spacetime
physics is discussed, in an approach quite distinct from the more restricted AdS-CFT
correspondence. A particular notion of holography in the context of black hole
thermodynamics is derived (rather than conjectured) from rather elementary
considerations, which also leads to a criterion of thermal stability of
radiant black holes, without resorting to specific classical
metrics. For black holes that obey this criterion, the canonical entropy is
expressed in terms of the microcanonical entropy of an Isolated Horizon which
is essentially a local generalization of the very global event horizon and  
is a null inner boundary of spacetime, with marginal outer trapping. It is argued why
degrees of freedom on this horizon must be described by a topological gauge
theory. Quantizing this boundary theory leads to the microcanonical
entropy of the horizon expressed in terms of an infinite series asymptotic in
the cross-sectional area, with the leading `area-law'  term followed by finite,
unambiguously calculable corrections arising from quantum spacetime fluctuations.   
\end{abstract}

\section{Introduction}

A black hole spacetime is characterized, as shown in Eddington-Finkelstein
coordinates for the Schwarzschild spacetime in Fig. 1. Contracting ellipses
depicting a gravitationally collapsing spherical star, are shown to form  a
marginally outer trapped
null surface called the Event Horizon. Local null cones are shown
to align with the EH, exhibiting the trapping behaviour characteristic of such
spacetimes. These tilt further inside the EH with a shrinkage showing
how the causal structure of spacetime is about to disappear at the
singularity shown as the dotted vertical line in the middle of the diagram,
which is the inevitable fate of the collapsing star and anything else entering
the EH.

\begin{figure}
\begin{center}
\psfig{file=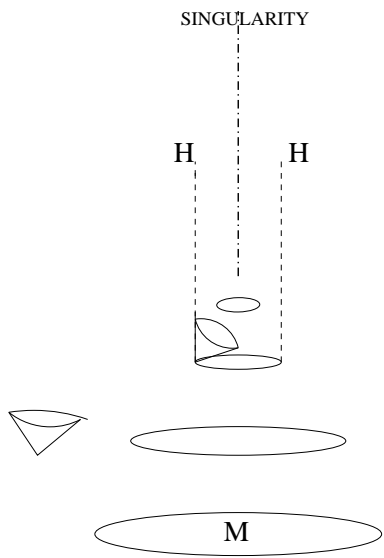,width=5cm}
\end{center}
\caption{Schwarzschild spacetime in Eddington coordinates}
\label{aba:fig1}
\end{figure}

An alternative view of the same spacetime is shown in Fig. 2 in conformal
coordinates which allows us to discuss asymptopia as a surface in a well-defined
part of spacetime. ${\cal I}^{\pm}$ are asymptotic future and past null
infinities for an asymptotically flat spacetime. A spherical collapse is
shown, together with the EH and also the singularity which clearly shows the
incompleteness of this spacetime vis-a-vis null and timelike geodesics
entering the EH. As shown in the figure, the black hole spacetime is defined
as the set of events of the universe ${\cal M}$ which do not lie in the
chronological past of ${\cal I}^+$, i.e., from which information (as light
signals say) never make it to future null infinity. This is the region ${\cal
B}$ shown in the figure. The EH is just the boundary of this spacetime region.

\begin{figure}
\begin{center}
\psfig{file=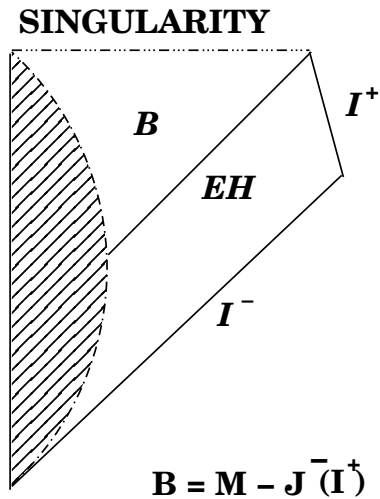,width=5cm}
\end{center}
\caption{Schwarzschild spacetime: conformal diagram}
\label{aba:fig2}
\end{figure}

As Subrahmanian Chandrasekhar puts it so eloquently in his treatise
Mathematical Theory of Black Holes,

\noindent {\it Black holes ... are the most perfect macroscopic
objects there are in the universe. The only elements in their
construction are our notions of space and time ... and because they
appear as ... family of exact solutions of Einstein's equation, they
are the simplest objects as well.} 

Yet black hole spacetimes have
\begin{itemize}
\item Singularities, where all known laws of physics break down.
\item Event horizon : boundary of spacetime accessible to asymptotic observers.
\end{itemize}
\noindent {\it  It is unlikely that black holes can be
understood on the basis of {\bf classical GR} even though their
horizons may have macroscopic cross-sectional areas !} 

Black holes have a further conundrum associated with the EH: the theorems on
Black Hole Mechanics \cite{bch72} derived from general relativity state that 
\begin{eqnarray*}
\delta {\cal A}_{EH} ~& \geq &~ 0 \cr
\kappa_{EH} ~&=&~ const \cr
\delta M ~&=&~ \kappa_{EH}~ \delta {\cal A}_{EH} + \cdots
\end{eqnarray*}
While these theorems exhibit an intriguing analogy with the laws of
thermodynamics, in reality there is no room for microstates in classical
general relativity for a family of exact solutions of Einstein's equation. 

In the early 1970s Bekenstein \cite{bek73} declared that {\bf Black holes
(must) have entropy}. The main argument is based on the Generalized Second Law
of Thermodynamics : $\delta(S_{out}~+~ S_{bh}) ~ \geq~ 0,$ where $S_{out}$ is
the entropy of all matter and radiaton outside the EH. Clearly, the existence of $S_{bh}$ is
essential for this law.  In order for this entropy to respect the second law
of black hole mechanics, independent of black hole parameters, it must be
proportional to the horizon area 
\begin{eqnarray}
S_{bh} ~=~ { {\cal A}_{hor} \over 4 l_P^2 }~ (k_B=1)~ \label{bhal}
\end{eqnarray}
\noindent where, $l_P \equiv (G \hbar /c^3)^{1/2} \sim 10^{-33} cm
\Rightarrow$ quantum gravity is necessary to provide the micro states whose
counting may eventually lead to black hole entropy. In any case, this implies that 
one needs to go beyond classical general relativity, in order to make sense of
entropy of black hole space times. Thus, black hole physics is by far the 
more compelling reason for quantizing spacetime geometry than
aesthetic reasons based on unification of fundamental interactions, of which
there is hardly any evidence observationally. 

Two issues that will have to be addressed imperatively before this idea can be
implemented:
\begin{itemize}
\item {\bf What degrees of freedom contribute to $S_{bh}$ ?}
\item {\bf How is it that $S_{bh}~=~S_{bh}(A_{hor})$ while
$S_{thermo}=S_{thermo}(vol)$ ?}
\end{itemize}

\section{Holography: a different approach}

A possible answer to the second question is provided by the so-called
{\bf Holographic Hypothesis} \cite{thf93}, \cite{bou02}, stated as follows \cite{thf93},

\noindent {\it ... Given any closed surface, we can
represent all that happens (gravitationally) inside it by degrees of
freedom on this
surface itself. This ... suggests that quantum gravity should be
described by a {\bf topological} quantum field theory in which all
(gravitational) degrees of freedom are projected onto the boundary.}
However, rather than use this as a working hypothesis, we adopt an alternative
viewpoint. We
\begin{itemize}
\item Propose : {\bf Holography is an outcome of the diffeomorphism invariance
of general relativity}. A version can be {\it derived} (heuristically).
\item We show: how gravitational degrees of freedom are projected to the
boundary for a particular model of the boundary known as a Isolated Horizon.
We also argue how these boundary degrees of freedom
are described by a three dimensional topological gauge theory on the
boundary, thus providing an explicit demonstration of a gravitation theory and
gauge theory connection. Once again, this is not a conjecture. 
\item Finally, we discuss important implications of this connection for $S_{bh}$.
\end{itemize}

\subsection{The proposal}

Diffeomorphism invariance $\Rightarrow$ {\bf there are no covariantly conserved
energy-momentum tensor for vacuum spacetimes in bulk in full nonlinear general
relativity}. Indeed, on the phase space of general relativity, diffeomorphism
generators appear as first class constraints. The Hamiltonian for bulk
spacetime is expressed as a linear combination of first class constraints,
\begin{eqnarray}
H_v &=& \int_{\cal S} \left[ N {\cal H} + {\bf N} \cdot {\bf P} \right] \\
&\approx& 0 ~{when}~ {\cal H} \approx 0,~ {\bf P} \approx 0 ~\label{ham}
\end{eqnarray}
where ${\cal H},{\bf P}$ are diffeomorphism generators and  $N (\rm{lapse}), {\bf N}
(\rm{shift})$ are Lagrange multipliers. In other words, there is no analogue
of ${\bf E}^2 + {\bf B}^2$  in vacuum general relativity in the bulk. Thus,
\begin{eqnarray}
H_{GR} ~=~ \underbrace{H_v}_{bulk} ~+~ \underbrace{H_b}_{boundary} ~. \label{b+bd}
\end{eqnarray}
On the constraint surface, $H_{GR} \approx H_b$, which implies that primary
excitations of quantum general relativity are not particle-like, 
but extended, like non-perturbative quantum chromodynamics. 

But, what about gravitons ? They are of course particle excitations of
perturbative quantum gravity, around weak gravitational backgrounds :
\begin{eqnarray}
g_{ab} ~=~ \underbrace{{\hat  g}_{ab}}_{fixed~ bkgd} ~+~
\underbrace{h_{ab}}_{graviton}
\end{eqnarray}
Thus, the description of gravitons requires
\begin{itemize}
\item a {\it fixed nondynamical} background
\item an expansion around a fiducial background, which is sensible only perturbatively
\item as such, it is quite inadequate for black hole thermodynamics.
\end{itemize}
In other words, black hole thermodynamics is {\it not} the thermodynamics of a
gas of gravitons in a non-dynamical gravitational background. It is rather the
thermodynamics of a black hole spacetime itself, i.e., of the geometry; this
is only possible if one can ascribe quantum states to spacetime geometry which
can be counted as microstates.  

\subsection{`Thermal' holography}

We now consider a canonical ensemble of radiant black hole spacetimes in
contact with a radiation bath at an inverse temperature $\beta$. The canonical
partition function is given by  
\begin{eqnarray}
Z(\beta) ~ = ~ Tr \exp -\beta {\hat H}~, ~\label{parti} 
\end{eqnarray}
where,
\begin{eqnarray}
{\hat H} ~ = ~ \underbrace{{\hat H}_v}_{blk}~+~\underbrace{
{\hat H}_{b}}_{bdy} ~. \label{hamil} 
\end{eqnarray}
The $Tr$ is over states defined as  
\begin{eqnarray}
|\Psi \rangle ~ = ~ \sum_{v,b} c_{vb} \underbrace{
 |\psi_v\rangle}_{blk} \underbrace{|\chi_b \rangle}_{bdy} \label{vxb}
\end{eqnarray}
i.e., the full Hilbert space ${\cal H}= {\cal H}_v \otimes {\cal H}_b$. The
{\bf Hamiltonian constraint} in the bulk implies that the quantum Hamiltonian
operator annihilates the bulk quantum states 
\begin{eqnarray}
{\hat H}_v ~|\psi_v\rangle ~=~ 0~. \label{hamilc}
\end{eqnarray}
It follows that
\begin{eqnarray}
Z(\beta) &=& \sum_{{}_{b}} \left( \sum_{{}_v} |c_{{}_{vb}}|^2 ||
  ~|\psi_{{}_v}\rangle~||^2
\right) \langle \chi_{{}_b}|\exp - \beta {\hat H}_{{}_{bdy}} |\chi_{{}_b}
\rangle \nonumber \\
&=& Tr_{{}_{bdy}} \exp -\beta {\hat H}_{{}_{bdy}} \nonumber \\
& \equiv & Z_{{}_{bdy}} ~. \label{bdyz} 
\end{eqnarray}
In other words, {\bf the bulk states decouple! } Boundary states determine  
bh thermodynamics completely : a thermal version of {\bf holography !} This is
different from the holographic hypothesis quoted above wherein {\it all} bulk
states are stipulated to be projected onto the boundary. 

\section{Isolated Horizons}

So far no specification of the kind of spacetime boundary we have in mind has been
made. Clearly, our interest is not in the asymptotic boundary. Instead we
focus on an {\it inner} boundary of spacetime. Recall that the event horizon
itself is a boundary of the chronological past of future asymptopia. But the
event horizon is too global for our purpose. It has the following lacunae:
\begin{itemize}
\item EH is {\it teleological} in nature, i.e., it is determined only after
{\it entire} spacetime is known. 
\item Stationarity $\Rightarrow$ black hole metric has a {\it global} timelike isometry.
\item Cosmological horizons (like the de Sitter horizon) cannot be
characterized as event horizons.
\item The (ADM) mass of the black hole is not defined on the event horizon but
as an integral over spatial infinity ($i^0$ in Fig. 2). 
\end{itemize}
In view of these shortcomings, we seek a {\it local} generalization of event
horizons. 

Such an alternative has already been found \cite{ash03} and is called an
Isolated Horizon (IQ). 

\begin{figure}
\begin{center}
\psfig{file=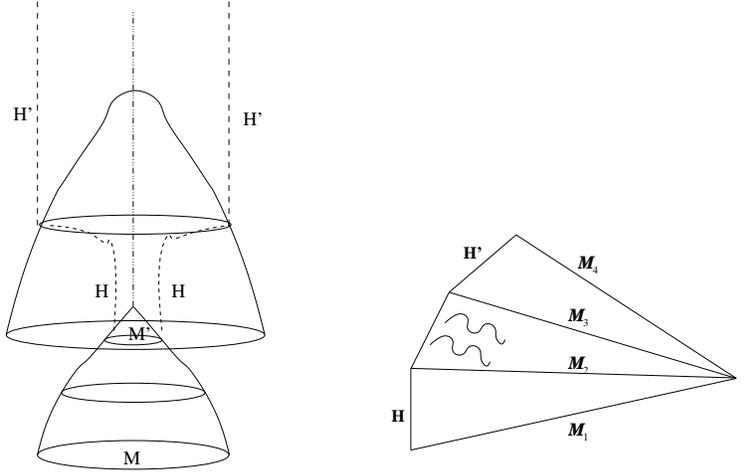,width=10cm}
\end{center}
\caption{Isolated Horizons}
\label{aba:fig3}
\end{figure}

We summarize the main properties of such a horizon, referring the reader to
\cite{ash03} for more details.
\begin{itemize}
\item An IQ has no global timelike isometry $\Rightarrow$ it is a {\bf
nonstationary} generalization of stationary event horizons, cosmological 
horizons, etc., allowing radiation to exist infitesimally close to it.
\item It is a null inner boundary of spacetime with topol ${\bf R} \otimes
{\bf S}^2$. 
\item The cross-sectional area ${\cal A}_{IH}$ of an IH remains constant: this is precisely
the {\it isolation}. Thus, nothing ever crosses an IH. 
\item {\it Zeroth law of IH Mechanics} The surface gravity $\kappa_{IH} = const$
\item On IH, can define mass $M_{IH} = M_{ADM} - {\cal E}_{rad}^{\infty} $ such
that $ \delta M_{IH} = \kappa_l \delta A_{hor} + \dots$ {\it (Ist law of IHM)}
\item Such horizons correspond thermodynamically to a  
{\it microcanonical ensemble with fixed $ {\cal A}_{IH} $ }. 
\end{itemize}

\subsection{Canonical entropy and thermal stability}

Consider now a canonical ensemble of IHs in contact with radiation; we proceed
to compute $S_{can}$ for this ensemble, assuming the equilibrium configuration
to be an IH with fixed $A_{IH}$ and $M_{IH} = M(A_{IH})$. Retaining Gaussian
fluctuations around a saddle point chosen to be this equilibrium configuration, we
get \cite{dmb01}, \cite{cm03} 
\begin{eqnarray}
S_{can} ~=~ S_{IH} ~+~ \frac12~\log S_{IH} ~\label{scan} 
\end{eqnarray} 
where $S_{IH}$ is the {\it microcanonical} entropy of the equilibrium IH.
Two issues arise immediately:
\begin{itemize}
\item What is $S_{IH}$ ? 
\item $S_{can} > 0 \Rightarrow$ black hole is thermally stable,
i.e., heat capacity $C > 0$. Under what conditions does this happen ?
\end{itemize}

We answer the second question first: the condition for thermal stability has
been determined \cite{cm05} \cite{pm07}
\begin{eqnarray}
{M_{IH} \over M_P} ~>~ {S_{IH} \over k_B}
\end{eqnarray}
This turns out to be the {\bf necessary and sufficient cond. for
$S_{can} > 0~ {\rm and} ~C > 0$}. Saturation of the inequality is seen to lead
to $C \nearrow \infty$ ! This is reminiscent of a first order phase
transition, even though here the transition is between {\it a stable and an
unstable phase}. This is similar to the Hawking-Page transition \cite{hawp83}
for an AdS-Schwarzschild black hole. The important distinction here is that 
it is completely general and also to an extent quantum in nature, in
contradistinction to the Hawking-Page treatment which is restricted to a
semiclassical analysis in anti-de Sitter black hole spacetime. It is seemingly 
generalizable to more general black holes with charge and angular
momentum, within the grand canonical ensemble. 

\subsection{IH as a null boundary: gravity-gauge link}

\noindent  Because IH is an {\it inner} boundary, we must add boundary term to the
action in order that the variational principle canbe used to derive Einstein's
equation. Thus,
\begin{eqnarray}
{\cal S} = {\cal S}_{EHL} + {\cal S}_{IH}
\end{eqnarray}
such that 
\begin{eqnarray}
\delta {\cal S}_{EHL}|_{IH} + \delta {\cal S}_{IH} ~ =~ 0
\end{eqnarray}
Since the IH is null the induced metric on it is degenerate
$\sqrt{^3g_{IH}}~=~0$. This has the consequence that the quantum theory
describing IH degrees of freedom must be a three dimensional topological field
theory for which the action is indep. of $^3g_{IH}$. Which 3 dim topological
field theory ? It must be a theory such that the degrees of freedom are
related in some manner to the bulk spacetime degrees of freedom (metric,
tetrad, connection). 

It turns out that with GR formulated in bulk as a gauge theory of the Poincare group (and
diffeomorphisms), the theory induced by the boundary conditions IH is an
$SU(2)$ Chern Simons gauge theory (in {\it time} gauge
where local Lorentz boosts are gauge fixed) with coupling constant $k \equiv
A_{IH} / 4\pi l^2_P >>> 1 $ \cite{ash97}. This provides one of the clearest
examples of a gravity-gauge theory link in the literature. This connection is
based on far firmer footing that others based on conjectured relationships.  
Using ${\cal S}={\cal S}_{{}_{EHL}} + S_{{}_{IH}}$, the variational principle
works, provided the following {\it consistency} condition holds
\begin{eqnarray}
\left(\frac{k}{2\pi} F_{CS} + E \times E  \right)_{{}_{S^2}} ~=~0~ . \label{cseom}
\end{eqnarray}
This is nothing but the Chern Simons theory equation of motion with the second
term functioning as source currents. This implies that the bulk spatial geometry characterized by
$ E $ plays the role of source for the Chern Simons degrees of freedom (given by $F_{CS}$)
characterizing boundary (IH) geometry. It is also a precise demonstration of
the {\it projection of bulk gravitational degrees of freedom to the boundary}
hypothesized in the Holographic Hypothesis. 

\section{Microcanonical entropy}

\subsection{Loop Quantum Gravity : spin network basis}

We now address the question of the microcanonical entropy of the IH
($S_{IH}$). The calculation follows the approach and methodology laid out in
\cite{ash97} - \cite{dkm01}. It is based on Loop
Quantum Gravity as well as the connection between Chern Simons gauge theories
and Wess-Zumino-Witten models \cite{wit89}. Loop Quantum
Gravity is perhaps the only known quantum theory of spacetime geometry, which
is background-independent and non-perturbative. It is a canonical version of
quantum general relativity, describing quantum three dimensional space (on
every spatial slice) in terms of {\it Spin Network} states. The Spin Network
basis was first proposed by Penrose and adapted to loop quantum
gravity by Rovelli and Smolin \cite{rov06}. Three dimensional space is
supposed to consist of fluctuating network graphs whose links each carry an $SU(2)$
irreducible representation index (`spin', $j=0,1/2,1,3/2,\dots$). Links meet
at vertices containing invariant $SU(2)$ `transporter' tensors constructed out
of the Levi-Civita tensor, depending on the valence of each vertex. An
arbitrary quantum state is a superposition of
spin network states which form an overcomplete basis.  

\begin{figure}
\begin{center}
\psfig{file=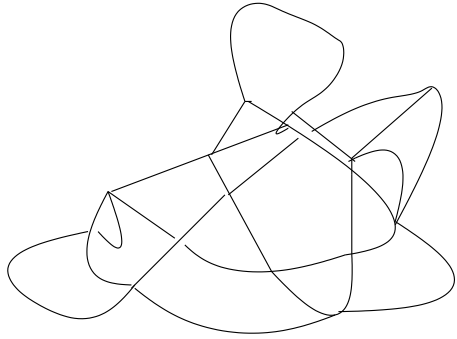,width=6cm}
\end{center}
\caption{Spin network graph}
\label{aba:fig4}
\end{figure}

The great advantage of the spin network basis is that geometric observables
(represented as self-adjoint operators), like length, area, volume, are
diagonal in this basis and turn out to have {\it discrete} spectra. In
particular, consider a spacelike two surface inserted into an arbitrary spin
network graph. The actual area of this surface will fluctuate around the
classical area $A_{cl}$ by terms $O(l_P^2)$ when the graph fluctuates, with
different spins puncturing the two-surface and transfering their spins to the punctures. 

The area operator, defined as  
\begin{eqnarray}
{\hat A}_S \equiv \sum_{I=1}^N \int_{S_I} {\det}^{1/2}[ ^2g({\hat E})] ~/label{arop}
\end{eqnarray}
can be shown \cite{rov06} to possess the bounded, discrete spectrum 
\begin{eqnarray}
a(j_1, \dots, j_N) &=& \frac14 \gamma l_P^2 \sum_{p=1}^N  \sqrt{j_p(j_p+1)} \\
\lim_{N \rightarrow \infty} a(j_1,....j_N) & \leq & A_{cl} + O(l_P^2) ~. \label{arsp} 
\end{eqnarray}

\subsection{ `Quantum' Isolated Horizon} 

\begin{figure}
\begin{center}
\psfig{file=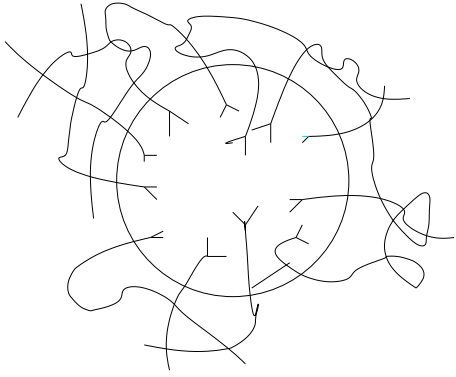,width=6cm}
\end{center}
\caption{Quantum isolated horizon}
\label{aba:fig5}
\end{figure}

Loop quantum gravity has not yet reached a stage of development where one can unambiguously
exhibit an IH formation from an appropriate solution of the quantum Einstein (Wheeler-de
Witt) equation, in some semiclassical approximation. Instead, we adopt an {\it
effective theory} viewpoint whereby we insert a foliation of the IH into the
spin network characterizing quantum spatial geometry, and use the formalism of
Chern Simons theory to obtain the states on this spherical section of the IH, 
with point sources carrying spin $j$ (arbitrary) on the punctures. Our
interest is to count $\dim {\cal H}_{CS + pt sources (j_1, ...j_n)}$ and get
$S_{IH}$ from 
\begin{eqnarray}
S_{IH}~\equiv ~ \log~\dim ~{\cal H}_{CS+(j_1,...,j_n)} ~, \label{sih}
\end{eqnarray}
for some fixed $A_{IH} >> l_P^2$, {\it restricting to only states with
vanishing total spin}. This latter restriction is enforced by the $SU(2)$
Gauss law constraint which implies that only rotationally invariant states are
physical. 

This computation is simplified by the
relation \cite{wit89} between the dimensionality of the CS theory Hilbert space and {\it
the conformal blocks of an $SU(2)_k$ WZW model living on the punctured
2-sphere}. Using this relation, and also the Verlinde formula, the
dimensionality of the Chern Simons Hilbert space is given by \cite{km98}
\begin{eqnarray}
\dim ~{\cal H}_{CS+(j_1,...,j_n)} &~=~& \prod_{p=1}^n \sum_{m_p=-j_p}^{j_p}
[\delta_{m_1 + \cdots + m_n,0} - \frac12 \delta_{m_1 + \cdots + m_n,-1} \\
&~-~& \frac12 \delta_{m_1 + \cdots + m_n,1}] ~. \label{kmf}
\end{eqnarray}
A moment's reflection on eq. (\ref{kmf}) is adequate to persuade us that
indeed the states with vanishing composite spin must have not only $m=0$ but
discounted by those states which have integral composite spin; the latter 
have not only an $m=0$ sector, but also $m=\pm1$ sectors. These nonvanishing
composite spin states do not satisfy the Gauss law constraint and have
to be eliminated if we are to consider only spinless states as
physical. Without this elimination, we have a larger degeneracy which will
ensue if the residual gauge invariance is $U(1)$ \cite{ash97} rather than
$SU(2)$. The reason we think it is natural to take $SU(2)$ rather than
$U(1)$ \cite{ash97} as the remnant of the local Lorentz invariance is that the former
is the invariance group relevant to the Gauss law constraint on the entire
spacetime, once Lorentz boosts are frozen out by choosing the `time' gauge. A
further gauge fixing to $U(1)$ on the IH \cite{ash97} appears to us to be
overly restrictive formally. Of course, one may desire to obtain the
degeneracy of the Chern Simons states for the entire Lorentz group as the
gauge group on the IH, but that task is made difficult by the fact that unitary
irreps of the Lorentz group are infinite dimensional. 

If, for simplicity we choose $j_p ~=~ \frac12~ \forall~ p=1, \dots, n$ we get
\begin{eqnarray}
S_{mc} ~=~ S_{IH} ~&=&~ \underbrace{{ A_{IH} \over 4 l_P^2}
}_{\rm Ashtekar ~et.al.~ 97} ~\\
&~-~&  \underbrace{\frac32 \log \left( { A_{IH} \over 4 l_P^2} \right) ~+~
const.~+~O(A_{IH}^{-1})}_{\rm Kaul~ \&~ PM~ 2000} ~. \label{kmf2}
\end{eqnarray}
The remarkable aspect of (\ref{kmf2}) is that, perhaps for the first time
since Bekenstein's pioneering work, one has an ab initio computation of
$S_{IH}$ and obtained an infinite series, asymptotic in $A_{IH}$, of quantum spacetime fluctuation
corrections to the Bekenstein-Hawking area law; each term of this series is
finite and unambiguously calculable. The leading correction to the area law is
logarithmic and has what appears to be a robust coefficient. With due modesty,
one may say that these corrections are the only known {\it physical}
signatures of loop quantum gravity as applied to the computation of {\it
microcanonical} black hole entropy.  

\section{Pending Issues}

\begin{itemize}
\item One needs to go beyond effective description in
terms of an embedded IH : we need to solve quantum dynamics and show the
formation of the horizon. 
\item We need to determine if Hawking radiation from IH is at all
possible, given its isolation.
\item We need to determine if the thermal nature of Hawking
radiation spectrum is an artifact of  the semiclassical approximation inherent
in the pioneering work. In other words, if a version of the horizon is shown
to radiate, then within a full quantum description, is the radiation of
quanta still in a thermal distribution ?
\item We need to understand if the lowest area quantum
$\sim l_{{}_P}^2$ has implications for the {\it information loss problem}.
\end{itemize}

\end{document}